\documentclass{webofc}
\usepackage[varg]{txfonts}
\usepackage{bm}
\usepackage[colorlinks, urlcolor=blue,linkcolor=blue, anchorcolor=blue, citecolor=blue]{hyperref}

\begin{document}
\title{Constraints on initial baryon stopping and equation of state from directed flow}

\author{\firstname{Lipei} \lastname{Du}\inst{1}\fnsep\thanks{\email{lipei.du@mail.mcgill.ca} (Speaker)} %
        \and
        \firstname{Chun} \lastname{Shen}\inst{2,3}\fnsep\thanks{\email{chunshen@wayne.edu}} %
        \and
        \firstname{Sangyong} \lastname{Jeon}\inst{1}\fnsep\thanks{\email{sangyong.jeon@mcgill.ca}} %
        \and
        \firstname{Charles} \lastname{Gale}\inst{1}\fnsep\thanks{\email{charles.gale@mcgill.ca}} %
}

\institute{Department of Physics, McGill University, Montreal, Quebec H3A 2T8, Canada 
            \and
           Department of Physics and Astronomy, Wayne State University, Detroit, Michigan 48201, USA 
            \and
           RIKEN BNL Research Center, Brookhaven National Laboratory, Upton, New York 11973, USA
}

\abstract{%
Our investigation focuses on the rapidity-dependent directed flow, $v_1(y)$, of identified hadrons in Au+Au collisions across a broad range of $\sqrt{s_{\rm NN}}$ from 7.7 to 200 GeV. Employing a (3+1)-dimensional hybrid framework, our study successfully reproduces the characteristic features of the measured $v_1(y)$ for both mesons and baryons across the considered beam energies. Notably, our analysis reveals the constraining power of baryonic $v_1(y)$ on the initial baryon stopping mechanism. Together with mesonic $v_1(y)$, the directed flow serves as a crucial tool for probing the equation of state governing dense nuclear matter at finite chemical potentials.
}
\maketitle
%
\section{Introduction}
\label{intro}

Exploration of the quantum chromodynamics (QCD) phase diagram heavily relies on heavy-ion collision experiments conducted at various beam energies \cite{Bzdak:2019pkr,Sorensen:2023zkk}. The intricate evolution of these collisions, spanning diverse phases, requires a multistage theoretical framework. That has successfully described numerous measurements. The collective flow of final hadrons offers crucial insight into early-stage dynamics, transport properties, and the equation of state (EoS) of the created dense nuclear matter \cite{Stoecker:2004qu}. The directed flow ($v_1$), signifying collective sideward motion, is especially sensitive to the early-stage evolution and the EoS \cite{Sorge:1996pc,Stoecker:2004qu}. The non-monotonic behavior of $dv_1/dy|_{y=0}$ (the slope of $v_1(y)$ near midrapidity) has been proposed as indicative of a first-order phase transition between hadronic matter and quark-gluon plasma (QGP) \cite{Brachmann:1999xt,Csernai:1999nf,Stoecker:2004qu}. This is because the softening of the EoS attributed to the phase transition can lead to a decrease in directed flow during expansion and consequently result in a minimum in $dv_1/dy|_{y=0}$ as a function of beam energy \cite{Stoecker:2004qu}. However, it's crucial to emphasize the sensitivity of $v_1(y)$ to diverse dynamical aspects. Various models have been utilized to compute $v_1(y)$ spanning from AGS to top RHIC energies, yielding widely varying results, and yet, none effectively describe the primary features of the measurements across the beam energies \cite{Singha:2016mna,Nara:2021fuu}. In this contribution, we explain the $v_1(y)$ for both mesons and baryons, using a (3+1)-dimensional hybrid framework with parametric initial conditions, and reveal its constraining power on the initial baryon stopping and the EoS of dense nuclear matter at finite chemical potentials \cite{Du:2022yok}.

\section{Setup and results}
\label{sec-1}

Our multistage framework initiates hydrodynamic evolution at a constant $\tau_0$, employing initial conditions derived from extending the nucleus thickness function with parametrized longitudinal profiles \cite{Denicol:2018wdp}. Within the hydrodynamic stage, we propagate the energy-momentum tensor and net baryon current, while accounting for dissipative effects from the shear stress tensor and net baryon diffusion current \cite{Denicol:2018wdp,Du:2019obx}. The framework uses the NEOS \cite{Monnai:2019hkn} that smoothly interpolates between lattice QCD EoS at high temperatures and a hadron resonance gas EoS at lower temperatures. As the system expands and cools, a transition to kinetic transport description occurs at a given local energy density. We utilize the measured pseudorapidity distribution of charged particle multiplicity, $dN^\mathrm{ch}/d\eta$, and the rapidity distribution of net-proton yields, $dN^{p-\bar{p}}/dy$, to constrain the longitudinal bulk dynamics \cite{Du:2022yok,Du:2023gnv}. The former provides stringent constraints on energy and entropy densities, while the latter on the baryon density.

The measured $dN^{p-\bar{p}}/dy$ exhibits two merging peaks across varying collision energies from 200 down to 5 GeV \cite{Videbaek:2009zy}. These two peaks are often attributed to incoming nucleons losing rapidity during interpenetration of the colliding nuclei (i.e., initial baryon stopping). A simple two-component profile, $f^B_-(\eta_s)$ and $f^B_+(\eta_s)$ (with $\eta_s$ being the spacetime rapidity), for initial baryon density to fit $dN^{p-\bar{p}}/dy$ results in an asymmetric baryon density distribution w.r.t.~the beam axis. After transverse expansion develops, this asymmetry would lead to a positively sloped $v_1(y)$ for baryons which significantly overshoots the observed data across all beam energies \cite{Voloshin:1996nv, Shen:2020jwv}. Reducing the initial baryon density near midrapidity can suppress baryon asymmetry and thus improve the simulated $v_1(y)$. However, this adjustment struggles to transport enough baryon charges to midrapidity to achieve the measured net proton yield due to the rapid decay of the baryon diffusion current \cite{Du:2021zqz}. Thus, despite posing significant challenges, aligning simulated $v_1(y)$ with measurements while reproducing $dN^{p-\bar{p}}/dy$ stands as an invaluable tool for discriminating between different models.

To tackle this challenge, we introduce an extra $\eta_s$-independent plateau component,  $f^B_c(\eta_s)$, within the baryon profiles, which has no preference for neither projectile nor target. This component contributes symmetrically to the baryon density w.r.t.~the beam axis, effectively suppressing the overall $v_1(y)$ of baryons. Meanwhile, sufficient net proton yields around midrapidity are achievable owing to the flat contribution of this plateau to the net proton yields across rapidity \cite{Du:2022yok}. Additionally, we incorporate an $\bm{x}_\perp$-dependent longitudinal shift in the baryon profiles to accommodate the local asymmetry in the thickness functions of projectile and target. This shift induces tilted peaks in the baryon density within the $x$-$\eta_s$ plane, accounting for the transverse variation in baryon stopping. The introduction of the plateau component in the baryon profile suggests a novel baryon stopping mechanism, potentially linked to the string junction conjecture \cite{Kharzeev:1996sq, Sjostrand:2002ip}. According to this conjecture, the baryon number within a nucleon correlates with the string junction, which is a non-perturbative Y-shaped configuration of gluon fields. 

\begin{figure}[tb]
\centering
\sidecaption
\includegraphics[width=.7\linewidth,clip]{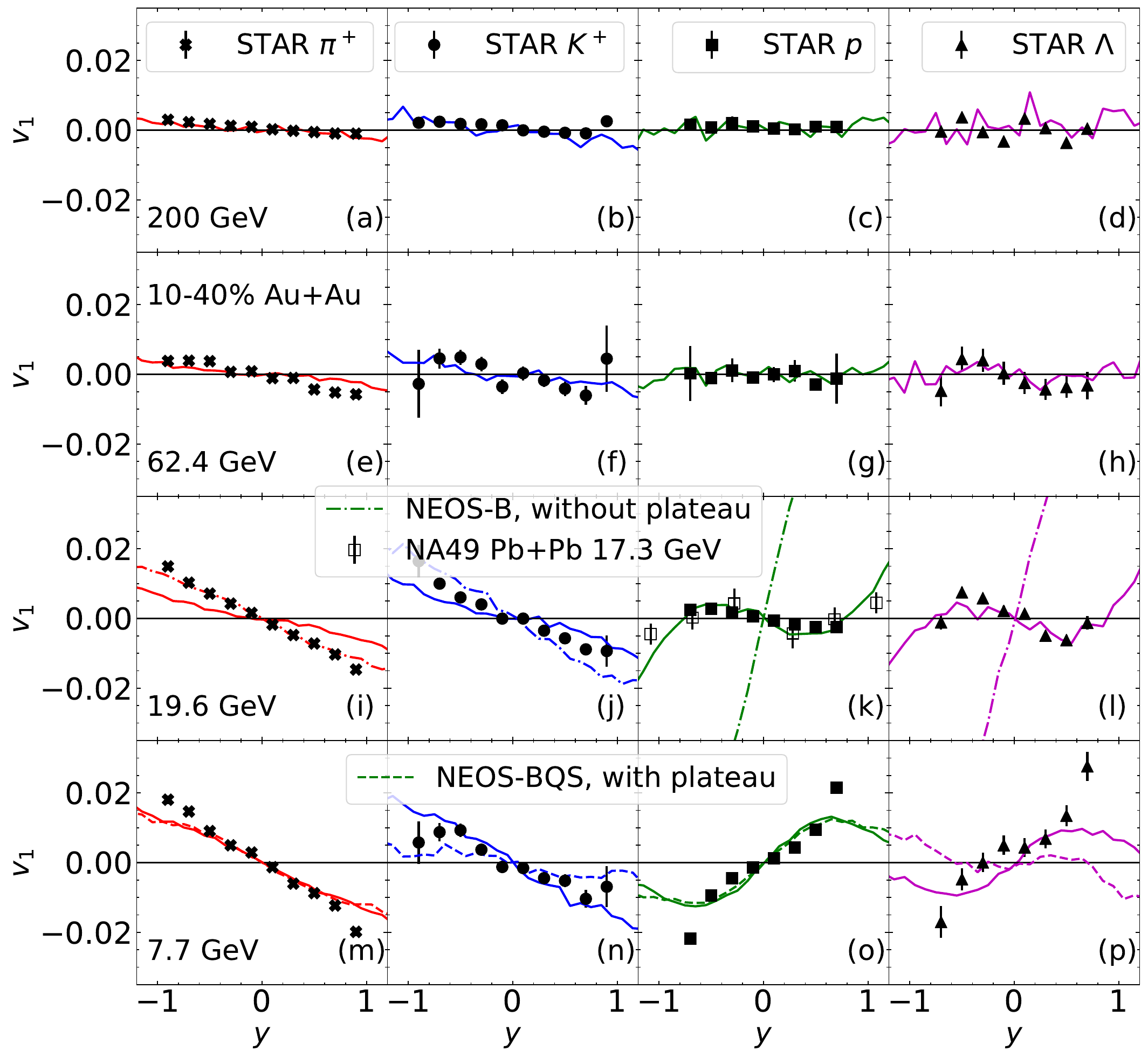}
\caption{%
Comparison between measured $v_1(y)$ (markers) and calculated values (lines) for $\pi^+$, $K^+$, $p$, and $\Lambda$ (from left to right) in 10-40\% Au+Au collisions across energies (7.7, 19.6, 62.4, and 200 GeV, bottom to top). The bottom two rows also illustrate $v_1(y)$ at 19.6 GeV using NEOS-B without the plateau (dot-dashed) and at 7.7 GeV with the plateau using NEOS-BQS (dashed), providing additional comparison. Figure is adapted from Ref.~\cite{Du:2022yok}.
}
\label{fig:bes_v1}
\end{figure}

Using the framework, we compute the $v_1(y)$ of four identified hadron species in 10-40\% Au+Au collisions using NEOS-B (assuming vanishing strangeness and electric charge chemical potentials, $\mu_S {\,=\,} \mu_Q {\,=\,} 0$)  \cite{Monnai:2019hkn}, demonstrating strong agreement with experimental measurements, as depicted by the solid lines in Fig.\,\ref{fig:bes_v1}. The $v_1(y)$ behavior of mesons predominantly arises from sideward pressure gradients, a well-established phenomenon in the field \cite{Bozek:2010bi}. What sets our model apart is its ability to reproduce the $v_1(y)$ patterns of $p$ and $\Lambda$ across the entire range from top RHIC energy down to 7.7 GeV, as illustrated in the right two columns of Fig.\,\ref{fig:bes_v1}. This success is largely attributed to the introduction of the plateau component within the initial baryon profile. Particularly at higher collision energies like 62.4 and 200 GeV, this plateau component dominates the midrapidity region, resulting in a nearly flat $v_1(y)$ for baryons within $|y|\lesssim1$. At 19.6 GeV, our model reproduces the distinctive features of the $v_1(y)$ pattern for baryons, including its cubic rapidity dependence (``wiggles''), when compared to STAR and NA49 measurements [see Fig.\,\ref{fig:bes_v1}(k,l)]. The baryon distribution with tilted peaks naturally accounts for the wiggles observed in $v_1(y)$ for baryons \cite{Du:2022yok}. Conversely, in the absence of $f^B_c(\eta_s)$, higher peaks in $f^B_-(\eta_s)$ and $f^B_+(\eta_s)$ would be required to reproduce the measured $dN^{p-\bar{p}}/dy$. This adjustment results in significantly stronger $v_1(y)$, exceeding the observed data. The dot-dashed lines in Figs.\,\ref{fig:bes_v1}(i-l) illustrate the model's recalibration to fit $dN^{p-\bar{p}}/dy$ without $f^B_c(\eta_s)$, consequently yielding a much stronger $v_1(y)$ for baryons, reaching a maximum value of $|v_1|_\mathrm{max}\approx 0.07$ around $|y|=1$, while the characteristic wiggles vanish.

At 7.7 GeV, the merging of the baryon peaks associated with $f^B_-(\eta_s)$ and $f^B_+(\eta_s)$ occurs due to the small beam rapidity, and as the transverse expansion develops, this baryon distribution naturally yields a $v_1(y)$ with a positive slope [see the solid lines in Fig.\,\ref{fig:bes_v1}(o,p)]. Notably, our obtained baryon $v_1(y)$ starts declining beyond $|y|\gtrsim0.7$, contrasting with the measured trend that continues to rise. This disparity may result from unaccounted interactions between spectators and the fireball, which are expected to be substantial at lower beam energies. We further investigate how the $v_1(y)$ of identified particles can reveal the nuclear matter chemistry by comparing results obtained using NEOS-B and NEOS-BQS (assuming strangeness neutrality $n_S{\,=\,}0$ and a fixed electric charge-to-baryon ratio $n_Q{\,=\,}0.4 n_B$) \cite{Monnai:2019hkn} at 7.7 GeV. Since NEOS-BQS requires local strangeness neutrality, it strongly correlates the production of $K^+$ and $\Lambda$, carrying the chemical potentials $\mu_Q+\mu_S$ and $\mu_B-\mu_S$, respectively. As shown by the dashed lines in Figs.\,\ref{fig:bes_v1}(m)-(p), this requirement disrupts the similarity indicated by measurements between $v_1$ of $\pi^+$ and $K^+$, as well as that between $p$ and $\Lambda$ at 7.7 GeV. This discrepancy suggests that to accurately describe the dynamical evolution around 10 GeV and lower center-of-mass energies, simulations considering multiple charge propagation with an EoS accounting for finite $\mu_{B,\,Q,\,S}$, unconstrained by NEOS-B and NEOS-BQS, would be necessary. Given the coverage of this low beam energy range by the second phase of Beam Energy Scan (BES-II) program, upcoming rapidity-dependent $v_1(y)$ measurements offer a means to probe the initial distributions and EoS at finite chemical potentials.

\section{Summary}
\label{summary}

Utilizing a (3+1)-dimensional hybrid framework with parametric initial conditions, we explain the observed rapidity and beam energy dependence of directed flow of identified particles in Au+Au collisions from 7.7 to 200 GeV. Introducing a rapidity-independent plateau in the initial baryon profile proves essential in explaining $v_1(y)$ features, especially the double sign changes at 19.6 GeV, while reproducing the rapidity-dependent hadron yields. Our findings indicate that the sign change in protons' $dv_1/dy|_{y=0}$ around 10-20 GeV is a consequence of initial baryon stopping, not evidence of a first-order phase transition. Additionally, our study suggests that directed flow measurements of strange particles around 10 GeV and lower energies could probe the EoS at finite chemical potentials. Particularly, our work highlights the significance of the plateau component, potentially linked to the string junction stopping conjecture. Implementing this component in more comprehensive frameworks and event-by-event simulations is crucial. Bayesian model selection, coupled with the forthcoming BES-II measurements, can help discern between different initial baryon stopping mechanisms, including scenarios involving string junctions.

\vspace{0.2cm}
\noindent {\textbf{Acknowledgments:}}
This work was supported in part (L.D., S.J., C.G.) by the Natural Sciences and Engineering Research Council of Canada, and in part (C.S.) by the U.S. Department of Energy, Office of Science, Office of Nuclear Physics, under DOE Award No. DE-SC0021969 and DE-SC0013460. C.S. also acknowledges support from a DOE Office of Science Early Career Award. 

\bibliography{refs}
\end{document}